\documentclass[preprint,12pt]{elsarticle}
\usepackage{graphicx}
\usepackage{amssymb}
\usepackage[T1]{fontenc}

\def\re{\mathop{\rm Re}\nolimits}
\def\im{\mathop{\rm Im}\nolimits}

\def\ln{\mathop{\rm ln}}

\newcommand{\rmd}{\ensuremath{\mathrm{d}}}
\newcommand{\rmi}{\ensuremath{\mathrm{i}}}
\newcommand{\rme}{\ensuremath{\mathrm{e}}}
\journal{Annals of Physics}
\begin{document}
\begin{frontmatter}

\title{Fine structure in the large $n$ limit of the non-hermitian Penner matrix model}

\author[ucm]{Gabriel \'Alvarez\corref{cor1}}
\ead{galvarez@fis.ucm.es}

\author[ucm]{Luis Mart\'{\i}nez Alonso}
\ead{luism@fis.ucm.es}

\cortext[cor1]{Corresponding author}

\author[uca]{Elena Medina}
\ead{elena.medina@uca.es}

\address[ucm]{Departamento de F\'{\i}sica Te\'orica II,
                               Facultad de Ciencias F\'{\i}sicas,
                               Universidad Complutense,
                               28040 Madrid, Spain}

\address[uca]{Departamento de Matem\'aticas,
                      Facultad de Ciencias,
                      Universidad de C\'adiz,
                      11510 Puerto Real, Spain}

\begin{abstract}
In this paper we apply results on the asymptotic zero distribution of
the Laguerre polynomials to discuss generalizations of the standard
large $n$ limit in the non-hermitian Penner matrix model. In these generalizations
$g_n n\to t$, but the product $g_n n$ is not necessarily fixed  to the value of the {}'t~Hooft coupling $t$.
If $t>1$ and the limit $l = \lim_{n\rightarrow \infty} |\sin(\pi/g_n)|^{1/n}$ exists, then the large $n$
limit is well-defined but depends both on $t$ and on $l$. This result implies that for $t>1$
the standard large $n$ limit with $g_n n=t$ fixed is not well-defined. 
The parameter $l$ determines a fine structure of
the asymptotic eigenvalue support: for $l\neq 0$ the support
consists of an interval on the real axis with charge fraction
$Q=1-1/t$ and an $l$-dependent oval around the origin with charge fraction $1/t$.
For $l=1$ these two components meet, and for $l=0$ the oval
collapses to the origin.  We also calculate the total electrostatic energy $\mathcal{E}$,
which turns out to be independent of $l$, and the free energy $\mathcal{F}=\mathcal{E}-Q\ln l$,
which does depend of the fine structure parameter $l$. The existence of  large $n$ asymptotic 
expansions of $\mathcal{F}$ beyond the planar limit  as well as the double-scaling limit are also discussed.
\end{abstract}


\end{frontmatter}
\section{Introduction}
The large $n$ limit with fixed {}'t~Hooft coupling $n 4\pi g_\mathrm{YM}^2=\lambda$ of non-abelian gauge 
theories has been the subject of intensive research for more than four decades, and in particular has fostered
the study of large $n$ matrix field theories.  Random matrix models are a simplified version of these theories
that offer a convenient setting to explore large $n$ limits, because the partition function of a $n\times n$
random matrix model can be written as an integral over the matrix eigenvalues $z_i$:
\begin{equation}
	\label{mm}
	Z_{n}(g)
	=
	\frac{1}{n!}\int_{\Gamma\times\cdots\times\Gamma}
	\prod_{j<k}(z_j-z_k)^2
	\exp\left(-\frac{1}{g}\sum_{i=1}^n W(z_i)\right)
	\prod_{i=1}^n \rmd z_i.
\end{equation}

The standard large $n$ limit is defined here as $n \rightarrow \infty$ with fixed {}'t~Hooft coupling $t= n g$.
This limit exhibits a number of interesting properties in the case of Hermitian matrix models with polynomial
potentials $W(z)$, such as the existence of an asymptotic eigenvalue density supported on a finite number
of real intervals (cuts), and the existence of a topological expansion of the free energy in the one-cut
case~\cite{BE79,BE80,JO88,ER03}.

\begin{figure}
    \begin{center}
        \includegraphics[width=8cm]{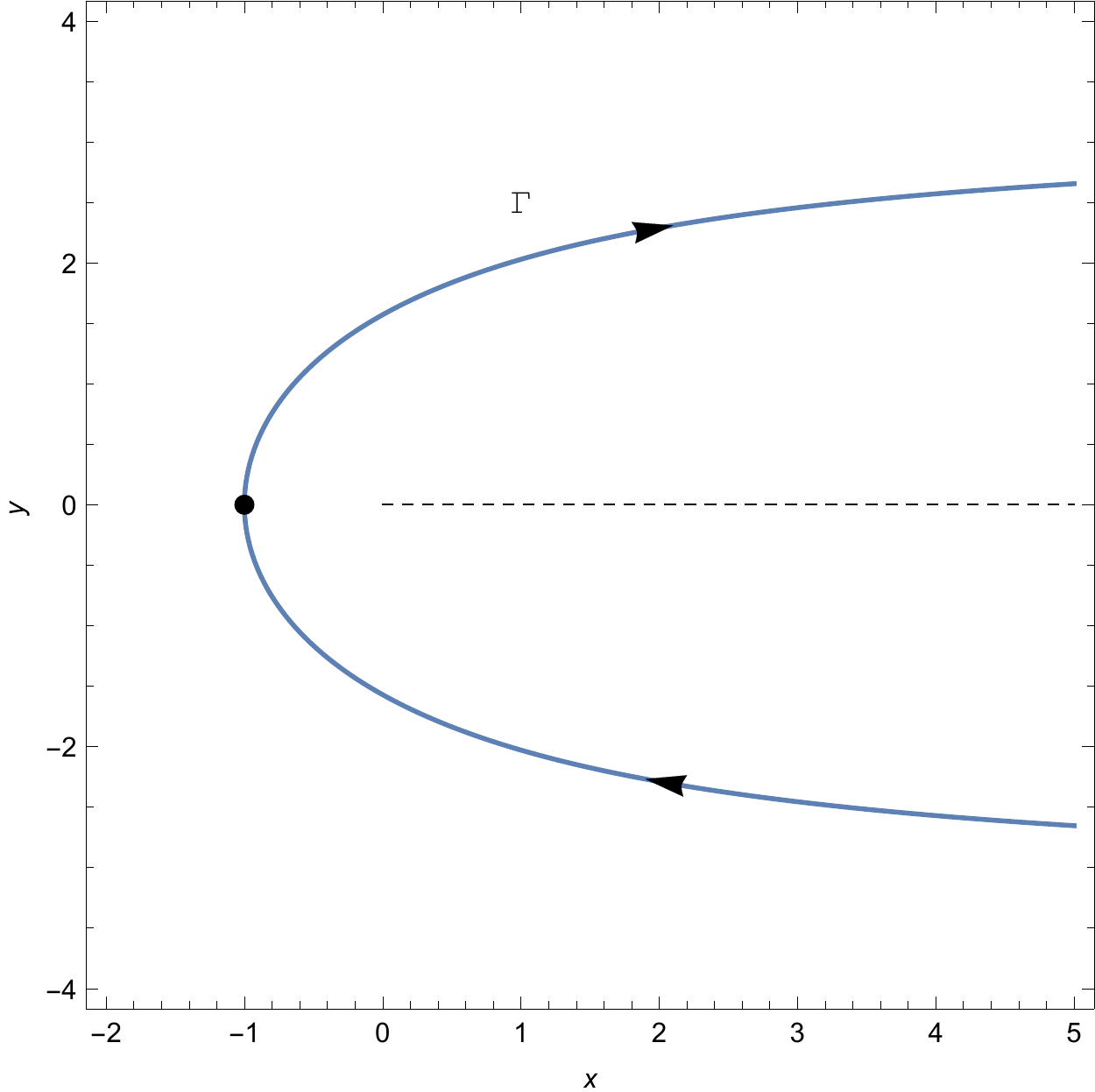}
    \end{center}
    \caption{Integration contour of the Penner model.\label{fig:path}}
\end{figure}

In this paper we discuss this large $n$ limit and generalizations thereof for the  Penner matrix model
with potential
\begin{equation}
	\label{pee}
 	W(z)= z+\log z.
\end{equation}
The logarithmic function in~(\ref{pee}) is defined by $\log z=\ln|z|+\rmi \arg z$ with $0\leq \arg z< 2\pi$
and where $\ln$ denotes the real logarithmic function on the positive real axis. We consider the Penner model on
any path $\Gamma$ in~(\ref{mm}) homotopic in $\mathbb{C} \setminus [0,+\infty)$ to the steepest
descent path $\im W(z)=\pi$ through the critical point $z=-1$  of $W(z)$ illustrated in~Fig.~\ref{fig:path}. 
Obviously, the corresponding partition function~(\ref{mm}) cannot be interpreted as an integral over the
(real) eigenvalues of a hermitian matrix model, but to a model in which the integration is performed
over a set of $n\times n$ complex matrices with $n$ eigenvalues constrained to lie on
the path $\Gamma$. In the literature~\cite{LA03,FE04} these models are frequently called
\emph{holomorphic} matrix models. However, to emphasize the relation of~(\ref{mm}) to the
theory of non-hermitian Laguerre orthogonal polynomials, hereafter we call our model the
non-hermitian Penner model.

The hermitian Penner model was introduced in~\cite{PE88} because the large $n$ expansion of its free energy
provided generating functionals for the  Euler characteristics $\chi_{k,s}$ of the moduli spaces of Riemann surfaces
of genus $k$ with $s\geq 1$ punctures. It was found later~\cite{DI90} that the analytic continuation of the free energy
in the hermitian case to the non-hermitian case and a suitable double scaling limit yielded a generating
functional for the Euler characteristics of unpunctured Riemann surfaces too. As a consequence, the Penner model
is closely related to the $c=1$ noncritical string with one direction of the spacetime
compactified on the circle with self-dual radius~\cite{DI90,CH91,TA91,TA92,AM94}.
More recently, the model has been used to analyze nonperturbative effects in $c=1$ string
theory and matrix models~\cite{MA06,PA10}.

In the early 90's Tan~\cite{TA91,TA92}, and Ambj{\o}rn, Kristjansen and Makeenko~\cite{AM94} applied
the saddle point method to give what was thought to be a complete description of the large $n$ limit of the Penner
model for both positive and negative fixed values of the {}'t~Hooft coupling.

The present paper is motivated by the behavior of the asymptotic zero distribution of the Laguerre polynomials
with negative parameter discovered by Kuijlaars and McLaughlin~\cite{KU04}. Note that fixing the {}'t~Hooft
coupling $t$ in effect defines a sequence
\begin{equation}
	\label{eq:thgn}
	g_n = \frac{t}{n}
\end{equation}
which trivially satisfies
\begin{equation}
	\label{ll}
	\lim_{n\rightarrow \infty} n g_n = t.
\end{equation}
We will show that the behavior discovered by Kuijlaars and McLaughlin~\cite{KU04} also arises in the
large $n$ limit of the non-hermitian Penner model if the sequence $g_n$ is not restricted to be~(\ref{eq:thgn}),
but only required to have the limit~(\ref{ll}) with $t>1$ and to have a finite value $l$ for the following limit:
\begin{equation}
	\label{eq:l}
 	l = \lim_{n\rightarrow \infty} |\sin(\pi/g_n)|^{1/n}.
\end{equation}

Under these conditions there indeed exists a large $n$ limit, but this limit depends not only on $t$
but also on the parameter $l$, introduced by Kuijlaars and McLaughlin in~\cite{KU04}. As a consequence,
we will also show that the standard large $n$ limit defined by the sequence~(\ref{eq:thgn}) with $t>1$
does not exist, i.e, in this limit there is not an asymptotic eigenvalue density with a well-defined support,
nor a planar limit of the free energy.
 
We begin our analysis by using the saddle point method~\cite{BE80,IT80,DI95,FE04} to establish the
connection between the asymptotic eigenvalue density of the Penner model and the asymptotic zero
density of the Laguerre polynomials. More concretely, we prove that given a sequence of coupling
constants $g_n$, the components of the (multiple) saddle points of the integrand of~(\ref{mm}) are the
zeros of the scaled Laguerre polynomials $L_n^{(-1-1/g_n)}(z/g_n)$. Therefore we can apply the results
of~\cite{KU04}, from which the following picture emerges: if $t>1$ and the limit $l$ exists, then there
exists an asymptotic eigenvalue density $\rho(z)$, and its support $\gamma_l$ consists of two pieces:
an interval $[a,b]$ on the positive real axis and an $l$-dependent closed loop $C_l$ around the origin.
These pieces are disjoint for $l\neq 1$. If  $l=0$ the loop reduces to the origin, where the asymptotic
eigenvalue density has a Dirac $\delta$ contribution.

In the electrostatic interpretation the eigenvalue density $\rho(z)$ represents a unit-normalized positive
charge density on $\gamma_l$ in the presence of an external electrostatic potential $V(z)/t$ given by
\begin{equation}
	\label{uve}
	V(z)	= \re W(z),
\end{equation}
and such that the total potential 
\begin{equation}
	\label{loge0}
	U(z) = \frac{1}{t}V(z) - 2 \int_{\gamma_l}  \ln |z-z'| \rho(z') |\rmd z'|
\end{equation}
is constant on each piece of $\gamma_l$. We calculate explicitly these constants, which turn out to be
different for $l\neq 1$. Therefore the density $\rho(z)$ for $l\neq 1$ is not an equilibrium density, but a
critical density in the sense of Mart\'{\i}nez-Finkelshtein and Rakhmanov~\cite{MA11}.
We also show that the total electrostatic energy
\begin{equation}
	\label{efe}
	\mathcal{E}=\frac{1}{t}\int_{\gamma_l}  V(z) \rho(z)|{\rm d}z|
	-
	\int_{\gamma_l} |\rmd z| \int_{\gamma_l} |\rmd z'| \ln |z-z'| \rho(z) \rho(z')
\end{equation}
is given by
\begin{equation}
	\label{e0}
 	\mathcal{E} =
	-\frac{1}{2}\ln t+\frac{3}{2}\Big({1-\frac{1}{t}}\Big) -\frac{1}{2}\Big({1-\frac{1}{t}}\Big)^2\ln(t-1).
\end{equation}
Note that  $\mathcal{E}$ is independent of $l$. 

We then apply the method of orthogonal polynomials~\cite{BE80,DI95} to derive an explicit expression of
the partition function~(\ref{mm}) of the non-hermitian Penner model in terms of the Barnes $G$ function, and then calculate the planar limit
of the free energy as
\begin{equation}
	\label{fee}
	\mathcal{F} = - \lim_{n\rightarrow \infty}\frac{\ln |Z_n(g_n)|}{n^2}.
\end{equation}
The main result here is that if $t>1$ and the limit $l$ exists, then
\begin{equation}
	\label{llz0}
 	\mathcal{F} = \Big(\frac{1}{t}-1 \Big) \ln l
				-\frac{1}{2}\ln t+\frac{3}{2}\Big(\frac{t-1}{t}\Big)
				- \frac{1}{2}\Big(\frac{t-1}{t}\Big)^2\ln (t-1),
\end{equation}
which involves an $l$-dependent term.  This proves that the $l$ dependence of the large $n$ limit
also manifests itself in the planar limit of the free energy.  We  illustrate this phenomenon with numerical
calculations.

Finally, by comparing~(\ref{e0}) and~(\ref{llz0}) it follows that the difference between the free energy
and the total energy can be written as
\begin{equation}
	\label{lar2}
	\mathcal{F} = \mathcal{E} - (\ln l) Q_{[a,b]}
\end{equation}
where
\begin{equation}
	\label{fff}
	Q_{[a,b]} = \int_a^b \rho(z) |\rmd z| = 1 - \frac{1}{t}
\end{equation}
is the amount of charge on the interval $[a,b]$. This is a somewhat unexpected
feature of the non-hermitian Penner model: for hermitian matrix models
with real analytic potentials $W(x)$ that grow faster than $\ln|x|$ at infinity
it is well known~\cite{JO88} that in the large $n$ limit
\begin{equation}
	\label{lar}
	\mathcal{F} = \mathcal{E}.
\end{equation} 

We also study in detail the expansion of the free energy beyond the planar limit for
the simple sequence $g_n=1/(n/t+\alpha)$, derive a suitable double scaling expansion
at the critical value $t=1$, and show that this latter expansion provides  generating functions for the Euler
characteristics of both unpunctured and punctured  Riemann surfaces.

In the last section of the paper we point out some extensions of its main results.
The technical details of the explicit computation of the total potential on the support
of the eigenvalue density are relegated to an appendix.
\section{Critical eigenvalue densities in the large $n$ limit\label{sec:sad}}
In this section we use the saddle-point method to derive the Schwinger-Dyson equation,
which is the main tool to determine the asymptotic eigenvalue density $\rho(z)$, and then
discuss the concepts of equilibrium and critical densities. The arguments are valid not only
for the Penner model~(\ref{pee}), but for any matrix model~(\ref{mm}) such that $W'(z)$
is a rational function of $z$ whose only singularities are at most a set of simple poles
$\mathcal{A}=\{a_1,\ldots,a_m\}$. We assume that $\Gamma$ is a path in the analyticity domain
of $W(z)$ such that the integral~(\ref{mm}) is convergent. The Penner model~(\ref{pee})
corresponds to a case in which $\mathcal{A}=\{0\}$.
\subsection{The saddle point method }
We  can write the partition function~(\ref{mm})  as 
\begin{equation}
	\label{mmm}
	Z_{n}(g_n)
	=
	\frac{1}{n!}	\int_{\Gamma \times\cdots\times\Gamma}
	\exp\Big(-n^2\mathcal{S}_n(z_1,\ldots,z_n)\Big)
	\prod_{i=1}^n \rmd z_i,
\end{equation}
where 
\begin{equation}
	\label{inte}
	\mathcal{S}_n = \frac{1}{g_n n^2}\sum_i W(z_i) - \frac{1}{2 n^2}\sum_{i}\sum_{ j\neq i}\log(z_i-z_j)^2.
\end{equation}

The saddle points ${\mathbf{z}}=(z_1,\ldots,z_n)$  of  $\mathcal{S}_n$ are the solutions of the equations
\begin{equation}
	\label{sa}
	\frac{1}{g_n}W'(z_i)+\sum_{j\neq i}\frac{2}{z_j-z_i}=0,
	\quad
	i=1,\ldots,n.
\end{equation}
Since these equations are  symmetric  under permutations  of the coordinates $z_i$, generically
each solution ${\mathbf{z}}=(z_1,\ldots,z_n)$ of~(\ref{sa}) gives rise to a set of $n!$  solutions
$\sigma({\mathbf{z}})=(z_{\sigma(1)},\ldots,z_{\sigma(n)})$, where $\sigma$ denote the permutations
of $n$ objects. 

The saddle point method assumes that there exists a sequence of saddle points
${\mathbf{z}}^{(n)}=(z_1^{(n)},\ldots,z_n^{(n)})$  of  $\mathcal{S}_n$ 
and a unit-normalized  positive density $\rho(z)$ (the eigenvalue density) with
support $\gamma$ such that
\begin{equation}
	\label{me}
	\frac{1}{n}\sum_{i=1}^n\delta(z-z_i^{(n)}) \rightarrow \rho(z) |\rmd z|,
\end{equation}
in the large $n$ limit (\ref{ll}). Each saddle point of the sequence ${\mathbf{z}}^{(n)}$
determines a monic polynomial which we denote by
\begin{equation}
	\label{fep}
	S_n(z) = \prod_i (z-z_i^{(n)}),
\end{equation}
and, as a consequence of~(\ref{sa}), the resolvent function 
\begin{equation}
	\label{re}
	\omega_n(z) = \frac{1}{n}\frac{S'_n(z)}{S_n(z)}
	                     = \frac{1}{n}\sum_{i} \frac{1}{z-z_i^{(n)}}
\end{equation} 
satisfies the Ricatti equation
\begin{equation}
	\label{ri}
	\frac{1}{n}\omega'_n(z) + \omega_n(z)^2-\frac{1}{n g_n}W'(z)\omega_n(z)
	=
	-\frac{1}{n^2 g_n}\sum_{i} \frac{W'(z)-W'(z_i^{(n)})}{z-z_i^{(n)}}.
\end{equation}
This equation coincides with the Ricatti equation~(2.6) derived in~\cite{DI95} after
performing the notational replacements $N\to n$, $g\to n g_n$ and $\omega \to -\omega_n$.

The large $n$ limit $\omega(z)$ of the resolvent $\omega_n(z)$ defined in~(\ref{re}) is
\begin{equation}
	\label{ef}
	\omega(z) = \int_{\gamma}\frac{\rho(z') |\rmd z'|}{z-z'}.
\end{equation}
Therefore, as a consequence of~(\ref{ri}), the function $\omega(z)$ must satisfy
the Schwinger-Dyson equation
\begin{equation}
 	\label{rl}
	\omega(z)^2-\frac{1}{t}W'(z)\omega(z)
	=
	-\frac{1}{t}\int_{\gamma}\frac{W'(z)-W'(z')}{z-z'}\rho(z') |\rmd z'|.
\end{equation}
For later reference, we also show the reduction to a second order linear equation
for $S_n(z)$ of the Ricatti equation~(\ref{ri}):
\begin{equation}
	\label{ri1}
	S''_n(z)-\frac{1}{ g_n}W'(z)\,S'_n(z)
	=
	-\frac{1}{ g_n}\left(\sum_{i} \frac{W'(z)-W'(z_i^{(n)})}{z-z_i^{(n)}}\right) S_n(z).
\end{equation}
\subsection{Critical densities} 
In this section we discuss briefly the minimum technical background that permits us to differentiate between
the equilibrium density (i.e, the absolute minimum) and critical densities in the sense of 
Mart\'{\i}nez-Finkelshtein and Rakhmanov~\cite{MA11}. In this context, it is convenient
to introduce the function
\begin{equation}
	\label{lai0}
	y(z) = \frac{1}{t} W'(z) - 2 \omega(z),
\end{equation}
and rewrite the Schwinger-Dyson equation~(\ref{rl}) as
\begin{equation}
	\label{lay}
	y(z)^2 = R(z),
\end{equation}
where
\begin{equation}
	\label{ere0}
	R(z) = \Big(\frac{1}{t}W'(z)\Big)^2-\frac{4}{t} \int_{\gamma}\frac{W'(z)-W'(z')}{z-z'}\rho(z') |\rmd z'|.
\end{equation}
Incidentally, the density $\rho(z)$ can be recovered from $y(z)$ using the Sokhotskii-Plemelj formulas.
Moreover, under our assumptions on $W'(z)$ the function $R(z)$ defined in~(\ref{ere0}) is a rational
function of $z$ with the same poles as $W'(z)$. Hence, from~(\ref{lai0}) it follows that the function
$y(z)$ is analytic outside $\gamma \cup \mathcal{A}$, and~(\ref{lay}) implies 
\begin{equation}
	\label{rh}
	y(z_+) = -y(z_-),\quad z\in \gamma.
\end{equation}

Equation~(\ref{lay}) is crucial to derive the main features of the eigenvalue density. 
Thus, we note that the  function $\omega(z)$  coincides with the Cauchy transform 
\begin{equation}
	\label{ct}
  	C^{\mu}(z) = \int_{\mathbb{C}}\frac{\rmd \mu(z')}{z-z'}
\end{equation}
of  the measure on $\mathbb{C}$, with support $\gamma$,  given by 
\begin{equation}
	\label{mea}
	\rmd \mu(z)=\rho(z)\, | \rmd z|.
\end{equation}
Hence~(\ref{lay}) can be written as
 \begin{equation}
 	\label{id}
 	\Big(\frac{1}{t} W'(z) - 2 C^{\mu}(z)\Big)^2 = R(z).
\end{equation}
This relation shows that $\rmd \mu(z)$ is a continuous critical measure on $\mathbb{C}$
in the sense of Mart\'{\i}nez-Finkelshtein and Rakhmanov~\cite{MA11}.  As a consequence
(see Lemma~5.2 of~\cite{MA11} and Proposition~3.8 of~\cite{KU14}), the support $\gamma$ of
$\rho(z)$ is a  union of a finite number of analytic arcs
\begin{equation}
	\label{cuts}
	\gamma = \gamma_1\cup\gamma_2\cup \cdots\cup \gamma_s,
\end{equation}
which are maximal trajectories of the quadratic differential 
\begin{equation}
	\label{qa}
	-y(z)^2 (\rmd z)^2 = -R(z)(\rmd z)^2,
\end{equation} 
i.e., maximal curves~\cite{st84} $z=z(t)\, (t\in (\alpha,\beta))$ such that
\begin{equation}
	-y(z)^2 \Big(\frac{\rmd z}{\rmd t}\Big)^2>0, \quad \mbox{for all $t\in (\alpha,\beta)$}.
\end{equation}
Moreover~\cite{MA11,KU14}, the total  potential 
\begin{equation}
	\label{loge}
	U(z)  = \frac{1}{t}V(z) - 2 \int_{\gamma}  \ln |z-z'| \rho(z') |\rmd z'| ,
\end{equation}
is locally constant on $\gamma$ 
\begin{equation}
	\label{s20}
	U(z)=u_i, \quad  z\in \gamma_i,\quad i=1,\ldots, s,
\end{equation}
with possibly different constants $u_i$, and that it satisfies 
\begin{equation}
	\label{s0}
	\frac{\partial U(z)}{\partial n_+} =  \frac{\partial U(z)}{\partial n_-},\quad z\in \gamma,
\end{equation}
where $n_{\pm}$ denote the two normal vectors to $\gamma$ at $z$ pointing in the opposite directions.
Equation~(\ref{s0}) is the so-called $S$-property of Stahl~\cite{ST85a,ST85b,ST86a}
and of Gonchar and Rakhmanov~\cite{GO84,GO89},  whose electrostatic interpretation is that
the electric fields at either side of $\gamma$ are opposite or, equivalently, that the forces acting on each
element of charge at $z$ from the two sides of $\gamma$ are equal.

The total electrostatic energy~(\ref{efe}) can be written in terms of the external potential and the constants $u_i$ as
\begin{eqnarray}
	\label{efeb}
	\mathcal{E} & = & \frac{1}{2t} \int_{\gamma} V(z) \rho(z) |\rmd z|
 	                      +
				 \frac{1}{2}  \int_{\gamma} U(z) \rho(z)  |\rmd z| \nonumber\\
                            & = & \frac{1}{2t} \int_{\gamma} V(z) \rho(z) |\rmd z|
                              +
                              \frac{1}{2} \sum_{i=1}^s u_i\,\int_{\gamma_i} \rho(z)  |\rmd z|.
\end{eqnarray}

We emphasize that unless all the constants $u_i$ are equal, the eigenvalue density is not an equilibrium
density minimizing the total energy~(\ref{efe}). In fact, continuous critical measures are characterized
by a different stationary condition for the total energy,
\begin{equation}
	\label{ec}
	D_h \mathcal{E}[\rho] = \lim_{s\rightarrow 0}\frac{\mathcal{E}[\rho^s]-\mathcal{E}[\rho]}{s}=0,
\end{equation}
under variations $\rho(z)\rightarrow \rho^s (z)$, with $\rmd \mu^s(z^s)=\rmd \mu(z)$,  induced by local set
transformations $z^s=z+s h(z)$ induced by smooth functions $h(z)$.

Summing up, in general the eigenvalue densities arising from the saddle point method determine
continuous critical measures on $\mathbb{C}$, and we will refer to them as critical densities.
\section{Laguerre polynomials and the asymptotic eigenvalue density  of the Penner model\label{sec:pe}}
In this section we specialize the general equations obtained in the previous section to the
Penner model~(\ref{pee}) and show its relation to the theory of Laguerre polynomials~\cite{DI90}
\begin{equation}
	\label{lagp}
	L^{(\alpha)}_n(z)=\sum_{k=0}^n\left(\begin{array}{c}n+\alpha \\n-k\end{array}\right)\frac{(-z)^k}{k!}.
\end{equation}

Indeed, the saddle point equations~(\ref{sa}) for the Penner model are
\begin{equation}
	\label{sa0}
	\frac{1}{g_n}\Big(1+\frac{1}{z_i^{(n)}}\Big)+\sum_{j\neq i}\frac{2}{z_j^{(n)}-z_i^{(n)}}=0,\quad i=1,\ldots,n,
\end{equation}
and the corresponding Ricatti equation~(\ref{ri}) is 
\begin{equation}
	\label{ripl}
	\frac{1}{n}\omega'_n(z)+\omega_n(z)^2-\frac{1}{n g_n}\Big(1+\frac{1}{z} \Big)\omega_n(z)
	=
	\frac{1}{n^2 g_n\,z}\sum_{i=1}^n \frac{1}{z_i^{(n)}}.
\end{equation}
From~(\ref{sa0}) it follows that
\begin{equation}
	\label{idpl}
	\sum_{i=1}^n \frac{1}{z_i^{(n)}}=-n,
\end{equation}
and we get the following  second order linear equation~(\ref{ri1}) for $S_n(z)$:
\begin{equation}
	\label{ri1b}
	S''_n(z)-\frac{1}{ g_n} \Big(1+\frac{1}{z} \Big)S'_n(z) = -\frac{n}{g_n\,z}S_n(z).
\end{equation}
By comparing~(\ref{ri1b}) with  the Laguerre differential equation
\begin{equation}
	\label{lage}
	z u''(z)+(\alpha+1-z) u'(z)+n u(z)=0,\quad u(z)=L^{(\alpha)}_n(z),
\end{equation}
we find that the monic polynomials $S_n(z)$ are proportional to the rescaled Laguerre polynomials
\begin{equation}
	\label{rel0} 
	L^{(\alpha_n)}_n\Big(\frac{z}{g_n}\Big),\quad \alpha_n=-1-\frac{1}{g_n}.
\end{equation}
Therefore, the saddle points  ${\mathbf{z}}^{(n)}=(z_1^{(n)},\ldots,z_n^{(n)})$
of the Penner model with coupling constants $g_n$ are given by
\begin{equation}
	\label{sopl}
	z_i^{(n)} = g_n\,l^{(\alpha_n,n)}_i,\quad i=1,\ldots,n,
\end{equation}
where  $l^{(\alpha,n)}_i$ are  the zeros of $L^{(\alpha)}_n(z)$.
\subsection{Zero asymptotics of scaled Laguerre polynomials}
Riemann-Hilbert and steepest-descent methods~\cite{KU04,MA01,KU01,DI11} have permitted
the complete characterization of the asymptotic zero distribution $\rho_L(z)$
of the scaled Laguerre polynomials  $L^{(\alpha_n)}_n(nz)$ for all real values of the parameter
\begin{equation}
	\label{clag}
	A = \lim_{n\rightarrow \infty} \frac{\alpha_n}{n}.
\end{equation}

We will concentrate on the case $-1< A<0$.   It turns out that the zeros cluster along certain curves
$\gamma$, but the form of $\gamma$ depends not only on the value of $A$ but also on the parameter 
  \begin{equation}
 	\label{ele}
 	l = \lim_{n\rightarrow \infty} |\sin(\alpha_n \pi)|^{1/n}
 	  = \lim_{n\rightarrow \infty} [\mathrm{dist}(\alpha_n, \mathbb{Z})]^{1/n},
\end{equation}
which represents the proximity degree of the sequence $\alpha_n$ to the integers. Thus, if the limit $l$ 
exists and we denote
\begin{equation}
	\label{end}
 	a_{\pm}=A+2\pm 2\sqrt{A+1},
 \end{equation}
then (see Theorem~1.2 and Eq.~(3.14) of~\cite{KU01})  it follows that ($\chi_{[a_-,a_+]}(x)$ equals 1 if
$x\in[a_-,a_+]$ and zero otherwise):
\begin{enumerate}
	\item For $l=0$ 
 		\begin{equation}
			\label{ge1}
			\gamma=\{0\}\cup [a_-,a_+],
		\end{equation}
		and
 		\begin{equation}
			\label{ge11}
  			\rho_L(z) |\rmd z| = -A \delta(z)+\frac{\sqrt{(x-a_-)(a_+-x)}}{2\pi x}\chi_{[a_-,a_+]}(x) \rmd x.
		\end{equation}
	\item For $l>0$ 
 		\begin{equation}
			\label{ge2}
			\gamma = \Gamma_l \cup [a_-,a_+],
		\end{equation}
 		where $\Gamma_l \subset \mathbb{C}\setminus (\{0\}\cup [a_-,+\infty))$ is a simple closed curve
		encircling $0$ clockwise, which is determined  by the implicit equation
		 \begin{equation}
		 	\label{cl}
 			\re \int_{a_-}^z \frac{\sqrt{(z'-a_-)(z'-a_+)}}{z'} \rmd z'=-\ln l.
 		\end{equation}
  		Furthermore
  		\begin{equation}
			\label{ge22}
    			\rho_L(z) |\rmd z| = \rmd\mu_l(z)+\frac{\sqrt{(x-a_-)(a_+-x)}}{2\pi x}\chi_{[a_-,a_+]}(x) \rmd x,
  		\end{equation}
  		where
  		\begin{equation}
			\label{dm}
  			\rmd\mu_l(z) = \frac{\sqrt{(z-a_-)(z-a_+)}}{2\pi\rmi  z} \rmd z,
  		\end{equation}
  		and  the branch of the squared root is such that $\sqrt{(z-a_-)(z-a_+)}\sim z$ as $z\rightarrow \infty$. 
\end{enumerate}
 
The case $l=1$ is generic  (see remark~1.3 in~\cite{KU04}) and it is the only value of $l$ for which
the loop $\Gamma_l$ and the interval $[a_-,a_+]$  intersect  (at  the point $a_-$).

We remark that this fine structure of the asymptotic zero distribution of the Laguerre polynomials does not arise
for other values of the limit~(\ref{clag}). Thus, if $-\infty<A<-1$ the support of the zero distribution is a well-defined simple analytic
curve symmetric with respect to the real axis~\cite{MA01}.
\subsection{The asymptotic eigenvalue density of the Penner model}
In view of~(\ref{rel0}) and~(\ref{sopl}), by setting
\begin{equation}
	\label{tra}
  	t=-\frac{1}{A}, \quad \gamma_l=\{tz,\, z\in\gamma\},\quad \rho(z)=\frac{1}{t}\rho_L\Big(\frac{z}{t}\Big),
\end{equation}
we can apply immediately the  results  on the asymptotic zero distribution of Laguerre polynomials 
into properties for the asymptotic eigenvalue density of  the Penner model.

Thus, it follows from (\ref{ge2})--(\ref{dm}) that for $t>1$ and $l> 0$ the support $\gamma_l$ of the  eigenvalue
density is given by 
\begin{equation}
	\label{nge2}
	\gamma_l=C_l \cup [a,b],
\end{equation}
where 
\begin{equation}
	\label{ylps2}
	a=2t-1- 2\sqrt{t(t-1)},\quad b=\frac{1}{a},
\end{equation}
and $C_l \subset \mathbb{C}\setminus (\{0\}\cup [a,+\infty))$ is the simple closed curve 
determined  by the implicit equation
\begin{equation}
	\label{cl2}
 	\re\int_{a}^z \frac{\sqrt{(z'-a)(z'-b)}}{z'}{\rm d}z' = -t \ln l.
 \end{equation}
Furthermore, the eigenvalue density is 
\begin{equation}
	\label{nge22}
    	\rho(z) |\rmd z|= \rmd\hat{\mu}_l(z) + \frac{\sqrt{(x-a)(b-x)}}{2\pi t x}\chi_{[a,b]}(x) \rmd x,
\end{equation}
where
\begin{equation}
	\label{dmn}
  	\rmd\hat{\mu}_l(z) = \frac{r(z)}{2\pi\rmi  t z}{\rmd } z,
 \end{equation}
with 
\begin{equation}
	\label{br}
   	r(z) = \sqrt{(z-a)(z-b)}
\end{equation}
such that $r(z)\sim z$ as $z\rightarrow \infty$. 

We will now show an important difference of these  large $n$ limits with respect to the hermitian case:
for $l\neq 1$ the eigenvalue density is not an equilibrium density but a critical density. Indeed,
let us write the total potential in the form
\begin{equation}
	\label{ug}
	U(z) = \frac{1}{t} V(z) - 2 \re g(z),
\end{equation}
where $g(z)$ is the function~\cite{KU04}
\begin{equation}
	\label{ge1b}
	g(z) = \int_{\gamma_l} \log(z-z') \rho(z') |\rmd z'|.
\end{equation}
For $z'\in[a,b]$ we take the cut of  $\log(z-z')$  along $[z',\infty)$,  while for $ z'\in C_l$ we take  the cut along
the part of $C_l$ that starts at $z'$, continues along $C_l$ in the clockwise sense until it reaches the real axis
and then runs to $+\infty$. Thus the function $g(z)$ is analytic in $\mathbb{C}\setminus (C_l\cup [a,\infty))$. 

Let $D_{\infty}$ denote the unbounded connected component of $\mathbb{C}\setminus (C_l\cup [a,\infty))$.
Then from~(\ref{nge22})--(\ref{br}) and using Cauchy's integration it follows that
\begin{equation}
	\label{ge2b}
	g'(z) = \frac{1}{2t}\Big( 1+\frac{1}{z}-\frac{r(z)}{z}\Big), \quad z\in D_{\infty},
\end{equation}
and as a consequence we have that for $z\in D_{\infty}$
\begin{equation}
	\label{ge3b}
	\re g(z) = \frac{1}{2t}\Big(x+\ln|z|-a-\ln a \Big)
	               -
	               \frac{1}{2t} \re\int_a^z\frac{r(z')}{z'} \rmd z'+  \re g(a).
\end{equation}
Hence from~(\ref{cl2}) and~(\ref{ug}) we obtain
\begin{equation}
	\label{ge4b}
	U(z)
	=
	\left\{\begin{array}{l}
		\displaystyle \frac{1}{t}(a+\ln a) - 2  \re g(a) - \ln l,
		 		    \quad \mbox{for $z\in\gamma_l$,} \\ \\
	        \displaystyle \frac{1}{t}(a+\ln a) - 2  \re g(a),
	                             \quad \mbox{for $z\in [a,b]$.}
	 \end{array}\right. 
 \end{equation} 
Therefore
\begin{equation}
	\label{une}
 	U\big |_{C_l} = U\big |_{[a,b]} - \ln l.  
\end{equation}
Since for $l\neq 1$ the constant values of the potential on each component of the support
are different, the eigenvalue density $\rho(z)$ is not an equilibrium density but a continuous critical density.
We defer to appendix~A the explicit evaluation of the constant value on the real interval, which turns out
to be
\begin{equation}
 	\label{val00}
	U\big |_{[a,b]} = \Big(2-\frac{1}{t} \Big) - \ln t - \Big(1-\frac{1}{t} \Big) \ln (t-1).
\end{equation}

\subsection{Existence of the eigenvalue density in the large $n$ limit\label{sec:el}}
It is worth noticing several striking consequences that the above results imply for the large $n$
limit~(\ref{ll}) of the Penner model.   We have seen that if $t>1$ and the limit $l$ exists, then
the asymptotic eigenvalue density of the Penner model exists but depends
not only on the value of the parameter $t$ but also on the value of $l$. We will prove now that
in the large $n$ limit with fixed {}'t~Hooft coupling $t>1$ there are subsequences
of the sequence $g_n=t/n$ that lead to different values of $l$ and therefore determine different
asymptotic eigenvalue densities. Consequently,  the large $n$ limit of the eigenvalue density
corresponding to the sequence $g_n=t/n$ does not exist.

We first observe that in terms of the sequence of coupling constants the parameter $l$ can be written as
\begin{equation}
	\label{elep}
 	l 
	  =  \lim_{n\rightarrow \infty}\Big |\sin\Big(\frac{\pi}{g_n}\Big)\Big|^{1/n}
	  = \lim_{n\rightarrow \infty} \Big |\sin\left(\pi\left\{\frac{1}{g_n}\right\}\right)\Big|^{1/n},
 \end{equation}
where $\{x\}$ stands for the fractional part of the real number $x$. 

Let us suppose that the fixed {}'t~Hooft coupling $t$ takes a rational value (in lowest terms)
\begin{equation}
	\label{rat} 
	t = \frac{p}{q},\quad  (p> q).
\end{equation}
Then $n q=j_n \, ({\rm mod}\, p) $ with  $j_n=0,1,\ldots,p-1$, and in the sequence of fractional parts
$\left\{ 1/g_n \right\} = \left\{ n q/p \right\}$ each number $0, 1/p,\ldots,(p-1)/p$,  appears infinitely often.
Therefore there are subsequences of $\left\{ 1/g_n \right\}$ with $l=0$ and subsequences with $l=1$.
Likewise, if  $t$ is irrational then the sequence  $\left\{ 1/g_n \right\}=\left\{ n/t \right\}$ is a dense subset
of the interval $[0,1]$. Hence $\left\{ 1/g_n \right\}$  has subsequences with $l=0$ and subsequences
with $l=1$.

There are simple examples of large $n$ limits~(\ref{ll}) such that the limit $l$ exists.
For instance, take $t$ rational of the form~(\ref{rat}) and define
\begin{equation}
	\label{e02}
	g_n = \frac{1}{n/t+\alpha},\quad \alpha>0.
\end{equation}
Decomposing again $n q=j_n \, ({\rm mod}\, p) $ with  $j_n=0,1,\ldots,p-1$, it follows that
the sequence $\left\{ 1/g_n \right\}$ takes the values $\left\{ \alpha\right\}, \left\{ \alpha+1/p\right\},\ldots\left\{ \alpha+(p-1)/p\right\}$.
Hence for real values of $\alpha>0$ such that
\begin{equation}
	\label{ejee}
	\alpha\neq \frac{k}{p},\quad k=1,2,\cdots
\end{equation}
the sequence $g_n$ yields the value $l=1$. 

Unfortunately, the example~(\ref{e02}) does not work for $t$ irrational. However, we may provide a more
general  example. Take the large $n$ limit with
\begin{equation}
	\label{e2}
	g_n = \frac{1}{[n/t]+c_n},
	\quad
	c_n = \frac{e^{-n r}}{2}, \quad n=1,2,\ldots
\end{equation}
where  $t>1$ and  $[x]$ denotes the integer part of $x$. We assume $0\leq r \leq + \infty$ and 
$e^{-\infty}=0$. Then it is obvious that $g_n n\rightarrow t$ and that
\begin{equation}
	\label{e13}
	l = \lim_{n\rightarrow \infty} c_n^{1/n}
	  = e^{-r}. 
\end{equation}
Therefore, all the possibilities~(\ref{ge1})--(\ref{dm}) arise.

\begin{figure}
    \begin{center}
        \includegraphics[width=8cm]{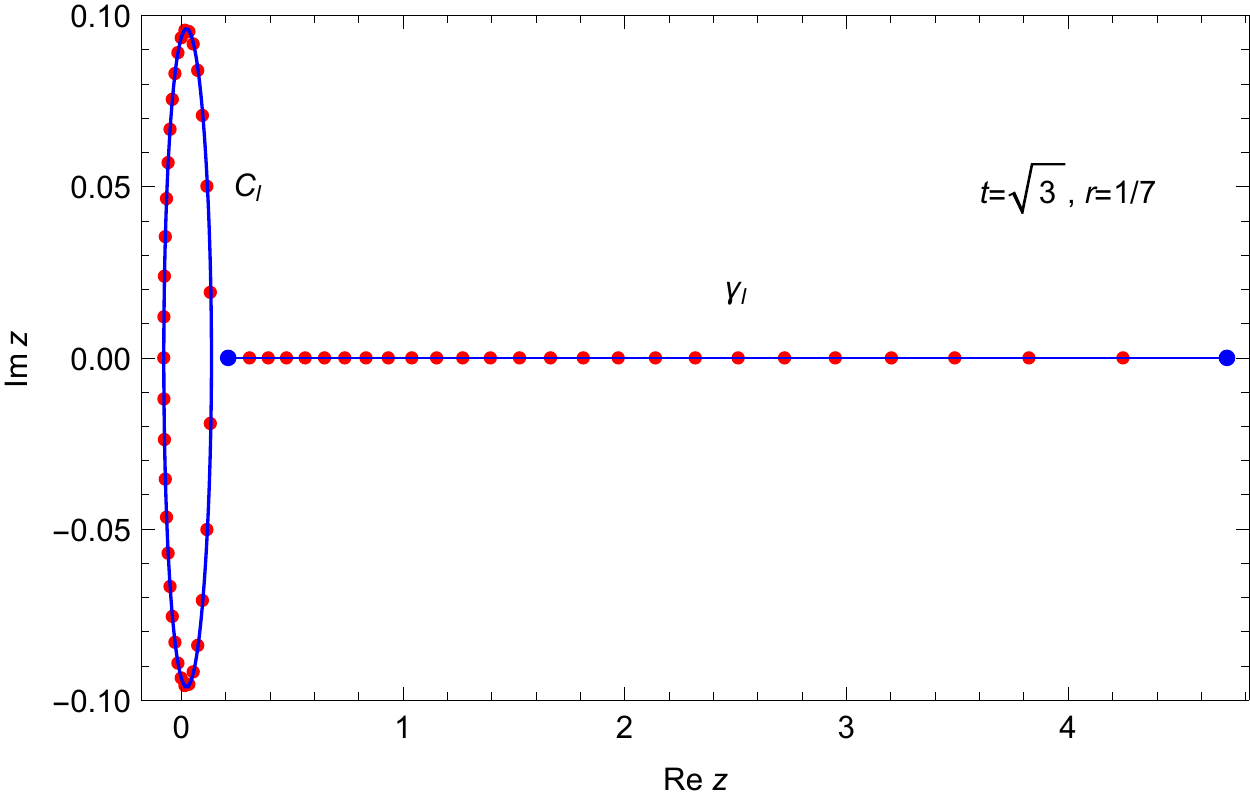}
    \end{center}
    \caption{Zeros of the $n=60$ scaled Laguerre polynomial and asymptotic density support
                  for a sequence $g_n$ of the form~(\ref{e2}) with $t=\sqrt{3}$ and $r=1/7$.\label{fig:supfig1v2}}
\end{figure}

\begin{figure}
    \begin{center}
        \includegraphics[width=8cm]{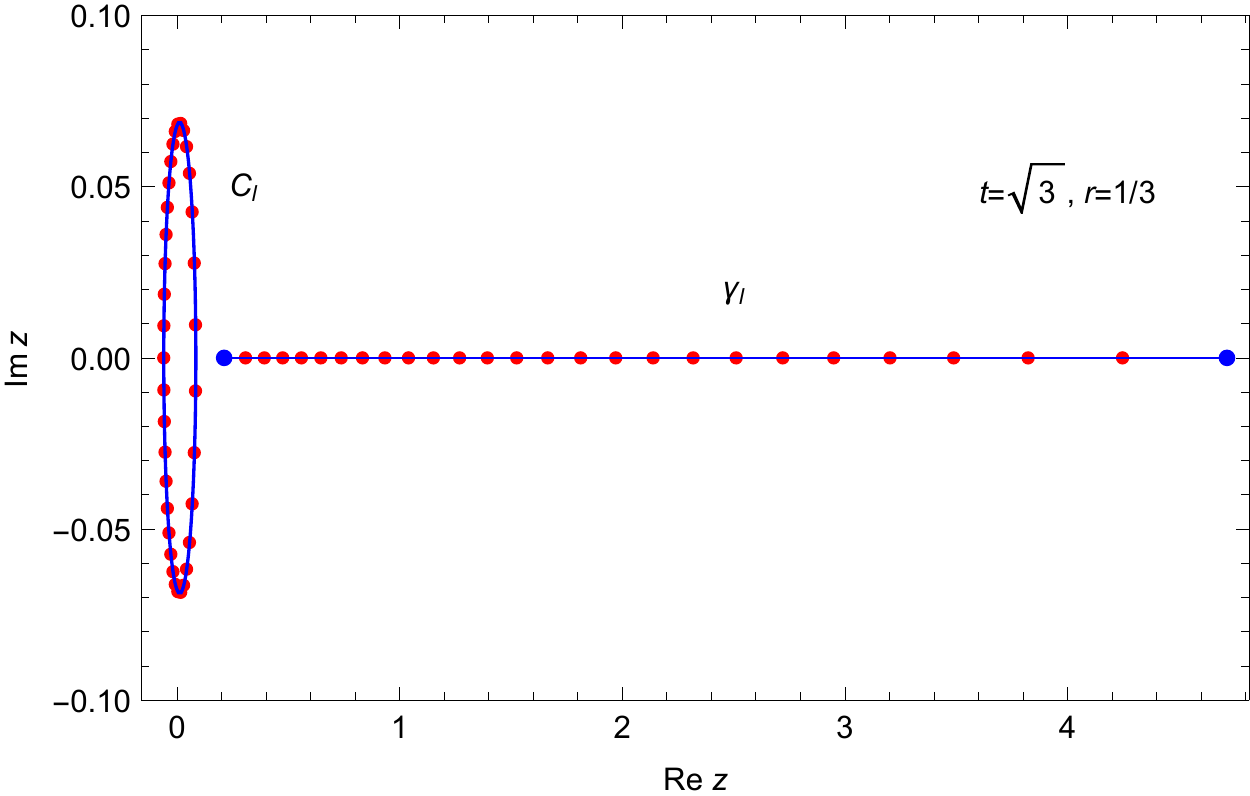}
    \end{center}
    \caption{Zeros of the $n=60$ scaled Laguerre polynomial and asymptotic density supported
                  for a sequence $g_n$ of the form~(\ref{e2}) with $t=\sqrt{3}$ and $r=1/3$.\label{fig:supfig2v2}}
\end{figure}

We may use this latter example to illustrate numerically the dependence of the support on $l$.
Figure~\ref{fig:supfig1v2} corresponds to a sequence~(\ref{e2}) with $t=\sqrt{3}$ and $r=1/7$.
The red markers denote the zeros of the corresponding scaled Laguerre polynomial with $n=60$,
which already accumulate distinctly on the two components $C_l$ and $[a,b]$ of the support,
which have been calculated numerically from~(\ref{ylps2}) and~(\ref{cl2}).
Figure~\ref{fig:supfig2v2} corresponds to a similar sequence~(\ref{e2}) with $t=\sqrt{3}$ but $r=1/3$.
Note how the endpoints~(\ref{ylps2}) $a\approx 0.21$ and $b\approx 4.72$ (marked in blue)
of the component of the support on the real axis are the same in both instances,
but the oval $C_l$ is notoriously different (both figures are drawn with the same scale).
Note also how in both cases 25 out of the 60 zeros are on the real interval, i.e.,
a fraction $25/60\approx 0.417$ already quite close to the limiting value
$1-1/t=1-1/\sqrt{3}\approx 0.423$.

Finally, we notice that  the  results of \cite{MA01} on the asymptotic zero distribution of Laguerre polynomials
for  values $-\infty<A<-1$  of the limit~(\ref{clag}) can also be applied to the non-hermitian Penner model.
These results imply the existence of an asymptotic eigenvalue density with a well-defined support
for any sequence $g_n$ of coupling constants with limit~(\ref{ll}) such that $0<t<1$.
\section{The  large $n$  limit  of the free energy }
The method of orthogonal polynomials~\cite{BE80,DI95}
\begin{equation}
	\label{pol1b}
	\int_{\Gamma} P_n(z) z^k \rme^{-\frac{1}{g} W(z)} \rmd  z=0,\quad k=0,\ldots,n-1,
\end{equation}  
leads to the following  expression  of  the partition function
\begin{equation}
	\label{zhr}
	Z_n = h_0^n \prod_{k=1}^{n-1} r_k^{n-k},
\end{equation}
where  
\begin{equation}
	\label{h0}
	h_0 = \int_{\Gamma} \rme^{-\frac{1}{g} W(z)}\rmd z,
\end{equation}
and $r_k$ are the recurrence coefficients in the three-term recursion relation
\begin{equation}
	\label{rec} 
	z P_k(z) = P_{k+1}(z) + s_k P_k(z) + r_k  P_{k-1}(z).
\end{equation}

For $\alpha\leq -1$ the  Laguerre polynomials  satisfy an orthogonality relation of the form~\cite{,KU04,KU01}
\begin{equation}
	\label{intl}
	\int_{\Gamma}  L^{(\alpha)}_n(z) z^k z^{\alpha} \rme^{-z} \rmd z=0,\quad k=0,1,\ldots,n-1,
\end{equation}
where $\Gamma$ is a path of the form shown in Fig.~\ref{fig:path} and $z^{\alpha}=\exp(\alpha\log z)$ 
with the branch of $\log$ defined in the Introduction. It is immediate that the rescaled Laguerre polynomials
\begin{equation}
	\label{as}
	L^{(-1/g)}_n\Big(\frac{z}{g}\Big)
\end{equation}
are proportional to the orthogonal polynomials $P_n(z)$~(\ref{pol1b}) corresponding to the Penner
model~(\ref{mm})--(\ref{pee}). We apply~(\ref{zhr})--(\ref{rec}) to calculate the partition function $Z_n$ of the Penner model.

Thus,  we have that
\begin{equation}
	\label{h0p}
	h_0 = \int_{\Gamma} z^{-\frac{1}{g} }\rme^{-\frac{z}{g}}\rmd z.
\end{equation}
This  integral  is an analytic  function of $g$ in $\mathbb{C} \setminus \{0\}$.  For $\re g>1$ we may
evaluate $h_0$ by approaching the contour $\Gamma$ to the positive real line  and then extend the result analytically.
Then, for positive values of $g$ we get
\begin{equation}
	\label{h2}
 	h_0 = \Big(1-\rme^{-\rmi\frac{2\pi}{g}}\Big) g^{1-\frac{1}{g}}\Gamma\Big( 1-\frac{1}{g}\Big).
 \end{equation}

Moreover, from the three-term recursion relation of the Laguerre polynomials
\begin{equation}
	\label{lag3}
	z L^{(\alpha)}_n(z)
	=
	-(n+1)L^{(\alpha)}_{n+1}(z)+(2n+\alpha+1) L^{(\alpha)}_n(z)-(n+\alpha)L^{(\alpha)}_{n-1}(z),
\end{equation}
the recurrence coefficients $r_k$ are given by 
\begin{equation}
	\label{recce} 
	r_k = k g (k g-1).
\end{equation}
 
At this point it is convenient to use the Barnes $G$ function~\cite{BA00,OL10} defined by
the canonical product
\begin{equation}
 	G(z+1) = (2\pi)^{z/2}
	               \rme^{-\frac{1}{2}(z+z^2(1+\gamma))}
	               \prod_{k=1}^\infty \left(1+\frac{z}{k}\right)^k \rme^{-z+z^2/2k},
\end{equation}
where $\gamma$ is the Euler-Mascheroni constant. This function satisfies
 \begin{equation}
	G(z+1)=\Gamma(z) G(z),\quad G(1)=1, 
\end{equation}
and therefore
\begin{equation}
	\label{id1}
	\prod_{k=1}^{n-1}(k+\alpha)^{n-k} = \frac{G(n+\alpha+1)}{G(\alpha+1) \Gamma(\alpha+1)^n}.
\end{equation}
Substituting~(\ref{h2}) and~(\ref{recce}) into~(\ref{zhr}) and using~(\ref{id1}) we find that the
partition function can be written as
\begin{equation}
	\label{znb}
	Z_n(g) = g^{n(n-\frac{1}{g})}\Big(1-\rme^{-\rmi\frac{2\pi}{g}}\Big)^n
	          \frac{G(n+1)G\Big(n+1-\frac{1}{g} \Big)}{G\Big(1-\frac{1}{g} \Big)}.
\end{equation}

The second factor in this expression is an essential  feature of the  class of  non-hermitian Penner models.
In order to  analyze the  large $n$ limit~(\ref{ll}) we write it as
\begin{equation}
	\label{ter}
	\Big(1-\rme^{-\rmi\frac{2\pi}{g}}\Big)^n
	=
	\rme^{-\rmi\frac{\pi}{g}n}(2\rmi)^n \sin^n\left(\frac{\pi}{g}\right).
\end{equation}

As to the third factor in~(\ref{znb}), when we evaluate $Z_n(g_n)$ for a sequence $g_n$ we need to evaluate the Barnes
$G$ functions for large positive and large (in absolute value) negative arguments. More concretely, if $t>1$
the terms $G(n+1)$ and $G(n+1-1/g_n)$ have to be evaluated for large positive values of their respective
arguments, while the term $G(1-1/g_n)$ in the denominator has to be evaluated for large (in absolute value)
negative values. However, if $0<t<1$ only the term $G(n+1)$ has to be evaluated for large positive
values, while both $G(n+1-1/g_n)$ and $G(1-1/g_n)$ have to be evaluated for large (in absolute value)
negative values.

The asymptotic expansion of $G(x)$ for large $x>0$ is well-known
and often written keeping an unexpanded gamma function
(see equation~5.17.5 in~\cite{OL10}), but we find more convenient the fully expanded version
\begin{eqnarray}
	\label{aba}
	\log G(x+1) & \sim &\frac{1}{2}x^2 \log x-\frac{3}{4}x^2+\frac{1}{2}x \log(2\pi)-\frac{1}{12}\log x+\zeta'(-1)
	\nonumber\\
			  &     &  {}+ \sum_{k= 2}^\infty\frac{B_{2k}}{2k(2k-2)}\frac{1}{x^{2k-2}}, \quad x\rightarrow\infty,
\end{eqnarray}
where $B_{2k}$ are the Bernoulli numbers and $\zeta'$ is the derivative of the Riemann zeta function.
Incidentally, the numbers $\chi_{k,0}=B_{2k}/(2k(2k-2))$ which appear in~(\ref{aba}) are precisely
the virtual Euler characteristics for the moduli space of unpunctured Riemann surfaces~\cite{HA86}.
To relate the values of the Barnes $G$ function on the negative axis to the values on the positive axis
we use the following expression~\cite{AD01},
\begin{equation}
	\label{eq:bneg}
 	G(-x) = (-1)^{\lfloor x/2\rfloor-1} G(x+2) \left|\frac{\sin(\pi x)}{\pi}\right|^{x+1}
	            \rme^{\frac{1}{2\pi}\mathrm{Cl}_2(2\pi(x-\lfloor x \rfloor))},
	            \quad x >0,
\end{equation}
where $\mathrm{Cl}_2(x)$ is the Clausen function (called Clausen's Integral and denoted by $f$ in~\cite{AS70}).
Therefore, for sequences $g_n$ such that the limit~(\ref{eq:l}) exists,
taking into account that the Clausen function is bounded, and that
\begin{equation}
	\label{je}
	\frac{\ln g_n}{g_n}=\frac{n}{t}\,\ln \Big(\frac{t}{n} \Big) +o(n),\quad n\rightarrow \infty,
\end{equation}
it follows  from (\ref{znb})--(\ref{eq:bneg}) that the free energy~(\ref{fee}) is
\begin{equation}
	\label{llz}
 	\mathcal{F} = H(t-1)\Big(\frac{1}{t}-1 \Big) \ln l
 				-\frac{1}{2}\ln t+\frac{3}{2}\Big(\frac{t-1}{t}\Big)
				- \frac{1}{2}\Big(\frac{t-1}{t}\Big)^2 \ln |t-1|,
\end{equation}
where $H(x)$ is the Heaviside step function.

\begin{figure}
    \begin{center}
        \includegraphics[width=8cm]{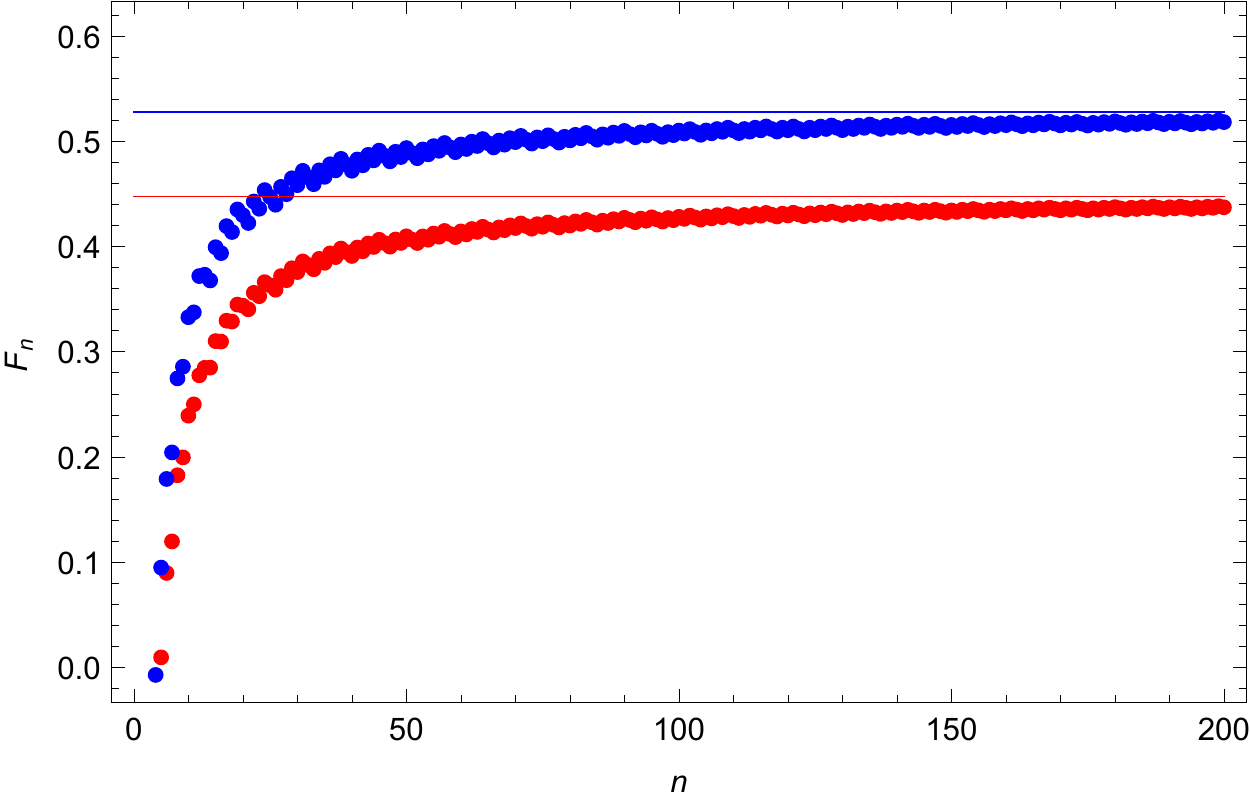}
    \end{center}
    \caption{Numerical values of $\mathcal{F}_n$ for a sequence with $t=\sqrt{3}$ and $r=1/3$
                 in~(\ref{e2}) (upper one, blue markers) and a similar sequence
                 with $t=\sqrt{3}$ and $r=1/7$ (lower one, red markers).
                 The horizontal lines are the respective limits given by~(\ref{llz}).\label{fig:tgt1}}
\end{figure}

As a numerical illustration of this result, in Fig.~\ref{fig:tgt1} we show the values of
\begin{equation}
	\label{eq:fn}
	\mathcal{F}_n = - \frac{\ln |Z_n(g_n)|}{n^2}
\end{equation}
for two sequences $g_n$: the upper one (blue markers) corresponds to $t=\sqrt{3}$ and $r=1/3$
in~(\ref{e2}), while the lower one (red markers) corresponds to $t=\sqrt{3}$ and $r=1/7$.
In this example $t>1$ and the horizontal lines are the respective (different) limiting values
given by~(\ref{llz}).

\begin{figure}
    \begin{center}
        \includegraphics[width=8cm]{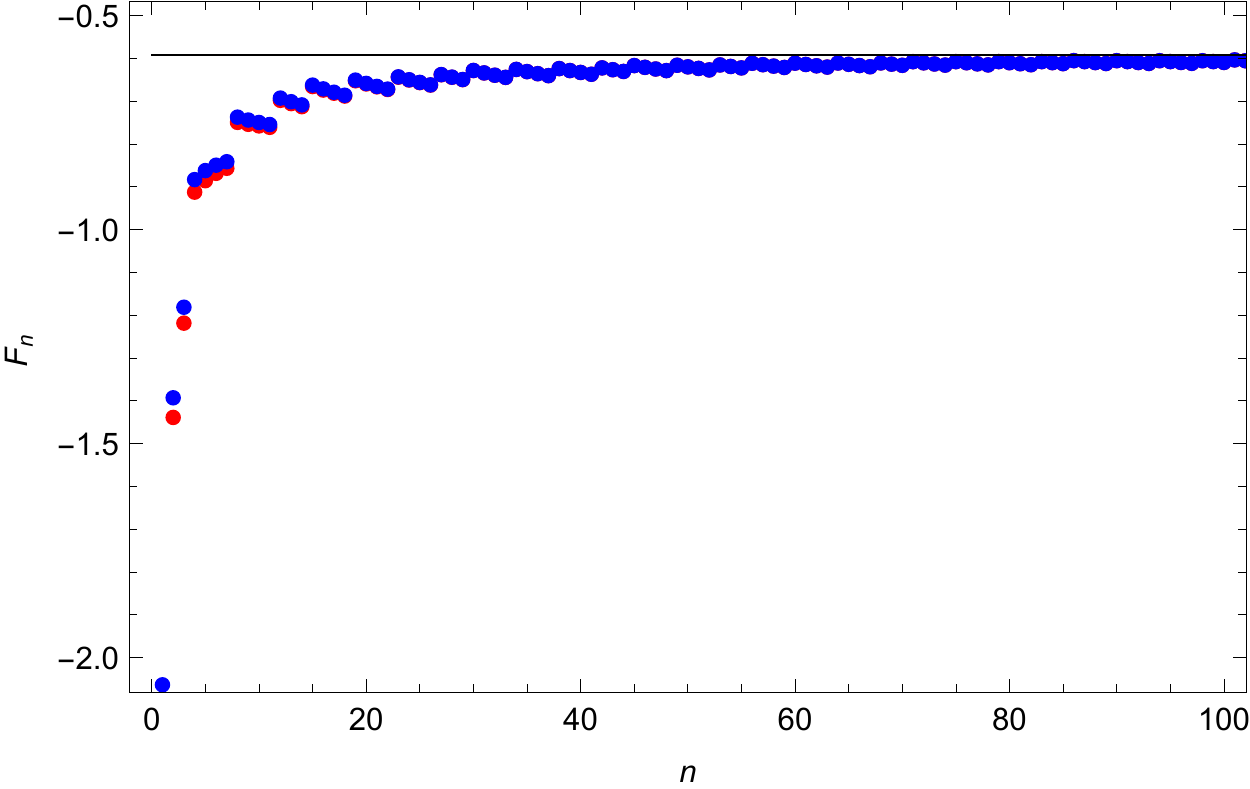}
    \end{center}
    \caption{Numerical values of $\mathcal{F}_n$ for a sequence with $t=1/\sqrt{3}$ and $r=1/3$
                 in~(\ref{e2}) (upper one, blue markers) and a similar sequence
                 with $t=1/\sqrt{3}$ and $r=1/7$ (lower one, red markers).
                 The horizontal line is the common limit given by~(\ref{llz}).\label{fig:tlt1}}
\end{figure}

Similarly, in Fig.~\ref{fig:tlt1} the upper, blue markers correspond
to $t=1/\sqrt{3}$ and $r=1/3$ in~(\ref{e2}), while the lower, red markers correspond to
$t=1/\sqrt{3}$ and $r=1/7$. In this case $t<1$ and the horizontal line is 
the common limit given by~(\ref{llz}).
\subsection{Free energy versus total energy}
For $t>1$  the expression~(\ref{llz}) for the planar limit of the free energy reads  
\begin{equation}
	\label{llz2}
	\mathcal{F}
	=
	\Big(\frac{1}{t}-1 \Big) \ln l -\frac{1}{2}\ln t+\frac{3}{2}\Big(\frac{t-1}{t}\Big)
	- \frac{1}{2}\Big(\frac{t-1}{t}\Big)^2\ln(t-1).
\end{equation}
We are going to prove that this identity can be written as
\begin{equation}
	\label{lar22}
	\mathcal{F} = \mathcal{E} - (\ln l) Q_{[a,b]}.
\end{equation}
where $Q_{[a,b]}$ is the charge contained in the interval $[a,b]$,
\begin{equation}
	Q_{[a,b]} = \int_a^b \rho(z) |\rmd z|,
\end{equation}
and  $\mathcal{E}$ is the total energy.

Using Cauchy's  theorem and~(\ref{dmn}) we obtain
\begin{equation}
	\label{in1}
	\int_{C_l}  \rho(z) |\rmd z| = -\frac{r(0)}{t} = \frac{1}{t}.
\end{equation}
Hence
\begin{equation}
	\label{in2}
	\int_a^b  \rho(z) |\rmd z| = 1-\frac{1}{t}.
\end{equation}
In this way the first term in~(\ref{llz2}) can be expressed as
\begin{equation}
	\label{in3}
	-(\ln l)\, Q_{[a,b]}.
\end{equation}

From (\ref{efeb}) and (\ref{une})  we have that the total energy is 
\begin{equation}
	\label{efe01}
	\mathcal{E} = \frac{1}{2t} \int_{\gamma_l}  V(z) \rho(z) |\rmd z|
	                      +
	                      \frac{1}{2}  U\big |_{[a,b]} - \frac{1}{2t} \ln l.
\end{equation}
The integral term in~(\ref{efe01}) can be also explicitly calculated (see appendix~A) 
\begin{equation}
	\label{intv0}
 	\int_{\gamma_l} V(z) \rho(z)  |\rmd z|
	=
	\ln l + t - 2 + \Big(1- \frac{1}{t}\Big)  \ln (t-1),
\end{equation}
and the final expression for the total energy takes the form
\begin{equation}
	\label{efin}
 	\mathcal{E}
	=
	-\frac{1}{2}\ln t+\frac{3}{2}\Big(\frac{t-1}{t}\Big) -\frac{1}{2}\Big(\frac{t-1}{t}\Big)^2\ln(t-1).
\end{equation} 
Therefore, the identity~(\ref{lar22}) follows.
It must be noticed  that, unlike the free energy, the total energy is independent on the value of $l$.
\subsection{Beyond the planar limit. The double scaling limit\label{sec:bpl}}
It has been rigorously proved for a large class of hermitian models with polynomial
potentials~\cite{BE79,BE80,ER03} that in the one-cut case the partition function
admits an asymptotic large $n$ expansion of the form
\begin{equation}
	\label{ae00}
 	-\ln \left|\frac{Z_n}{Z_n^{\mathrm{G}}}\right|
	\sim
	\sum_{k=0}^{\infty} F_{k}\, n^{2-2k},
\end{equation}
where $Z_n^{\mathrm{G}}$ denotes the partition function of the Gaussian model $W(z)=z^2/2$.
It is also well-known the appearance of oscillatory terms in the multi-cut case~\cite{BO00}.
The question naturally arises if similar expansions can be generated for the non-hermitian
Penner model from~(\ref{znb}) and suitable sequences $g_n$ satisfying~(\ref{ll}).

Let us consider the one-parameter family of sequence of coupling constants
\begin{equation}
	\label{e023}
	g_n = \frac{1}{n/t+\alpha},\quad \alpha \geq 0.
\end{equation}
As we have seen above in Sec.~\ref{sec:el}, if $t$ is a rational number 
$t= p/q>1$  and $\alpha\neq k/p \; (k=0,1,2,\ldots) $  then the sequence $g_n$ determines
a well defined large $n$ limit with $l=1$.   
 Substituting this sequence $g_n$ in the exact
equation for the partition function~(\ref{znb}) and expanding in $n$ we obtain
\begin{equation}
	\label{ae}
	-\ln |Z_n| \sim\frac{1}{12}\ln n+ \sum_{k=0}^{\infty} n^{2-k}\,F_k,
\end{equation}
where the coefficients $F_n$ are given by:
\begin{equation}
	F_0
	 =
	-\frac{1}{2}\ln t +\frac{3}{2}\Big(\frac{t-1}{t}\Big)-\frac{1}{2}\Big(\frac{t-1}{t}\Big)^2\ln(t-1),
\end{equation}
\begin{eqnarray}
	F_1
	& = & -\ln(2\pi) - \alpha\left(2-t-\frac{t-1}{t}\ln(t-1)\right) \nonumber\\
	&    & {}-\frac{t-1}{t}\ln\left|2\sin\left(\pi\left(\frac{j_n}{p}+\alpha\right)\right)\right|,
\end{eqnarray}
\begin{eqnarray}
	F_2
	& = & -\zeta'(-1) + \frac{1}{12}\ln(t-1) - \frac{\alpha^2}{2}\left(t(t+1)+\ln(t-1)\right)\nonumber\\
	&    &  {}+ \alpha \ln \left| 2\sin\left(\pi\left(\frac{j_n}{p}+\alpha\right)\right)\right|
	             + \frac{1}{2\pi}{\mbox{Cl}}_2\left(2\pi\left(\frac{j_n}{p}+\alpha\right)\right),
\end{eqnarray}
\begin{eqnarray}
	F_k
	& = &
	(-1)^{k+1} \alpha^kt^{k-1}\frac{(k-1)t+1}{k(k-1)}
	- \frac{B_k}{k(k-2)} \nonumber\\
	&   &
	{}- t^{k-2}\left(\frac{1}{(t-1)^{k-2}}-(-1)^k\right)\times\nonumber\\
	&  &
	\left(\sum_{j=2}^{[\frac{k}{2}]}\frac{B_{2j}}{2j(2j-2)}
       \left({ k-3 \atop 2j-3 }\right)\alpha^{k-2j}-\frac{\alpha^k}{k(k-1)(k-2)}+\frac{\alpha^{k-2}}{12(k-2)}\right),
       \nonumber\\
       &  & \qquad\qquad\qquad\qquad\qquad\qquad
               \qquad\qquad\qquad\qquad (k\geq 3),
\end{eqnarray}
and $j_n=n q \, (\mathrm{mod}\, p)$. Note that the expression for $F_0$ coincides with
the $\mathcal{F}$ of~(\ref{llz}) for $t>1$ and $l=1$. Note also that this expansion~(\ref{ae})
features oscillatory terms in the coefficients $F_1$ and $F_2$. 
The coefficients $F_k$  with $k\geq 2$ diverge as $t\rightarrow 1+$, so that the limit value $t=1$ is critical.

In order to study the free energy near the critical value $t=1$  and to avoid the limitation of using rational values
of $t>1$,  we consider the sequence~(\ref{e023}) for $0<t<1$ and arbitrary values of $\alpha\geq 0$.
In this case, substituting the sequence $g_n$ in~(\ref{znb}) and expanding in $n$ we obtain an asymptotic
expansion of the form~(\ref{ae})  with the same coefficients $F_k$ for $k\geq 3$ but
\begin{equation}
	F_0
	 =
	-\frac{1}{2}\ln t +\frac{3}{2}\Big(\frac{t-1}{t}\Big)-\frac{1}{2}\Big(\frac{t-1}{t}\Big)^2\ln(1-t),
\end{equation}
\begin{equation}
	F_1= -\ln(2\pi) - \alpha\left(2-t-\frac{t-1}{t}\ln(1-t)\right),
	\end{equation}
\begin{equation}
	F_2
=-\zeta'(-1) + \frac{1}{12}\ln(1-t) - \frac{\alpha^2}{2}\left(t(t+1)+\ln(1-t)\right).
\end{equation}
Note that the oscillatory terms arising in the expansion for $t>1$ have disappeared.
At $t=1$ the coefficient $F_2$ has a logarithmic singularity  while the  coefficients $F_k$  with $k\geq 3$
have a pole of increasing order:
\begin{eqnarray}
		 F_k & \sim & -\frac{1}{(t-1)^{k-2}}
	                       \Bigg(\sum_{j=2}^{[\frac{k}{2}]}\frac{B_{2j}}{2j(2j-2)}
                               \left({ k-3 \atop 2j-3 }\right)\alpha^{k-2j} \nonumber\\
                 &  & {}- \frac{\alpha^k}{k(k-1)(k-2)}
                                         +\frac{\alpha^{k-2}}{12(k-2)}\Bigg), \quad k\geq 3.
\end{eqnarray}
The sum of the divergent  terms with $k\geq 3$ (as $t\to 1+$)  of the asymptotic expansion of the free energy 
can be formally written as
\begin{eqnarray}
	\label{sing}
	-\ln \left| Z_n^{\mathrm{sing}} \right|
	& \sim &
	-\sum_{k=3}^{\infty} \Big(n(t-1)\Big)^{2-k} \sum_{j=2}^{[\frac{k}{2}]}\frac{B_{2j}}{2j(2j-2)}
                                        \left({ k-3 \atop 2j-3 }\right)\alpha^{k-2j} \nonumber\\
	&        & {}+ \sum_{k=3}^{\infty}
	                   (n(t-1))^{2-k} \left(\frac{\alpha^k}{k(k-1)(k-2)}- \frac{\alpha^{k-2}}{12(k-2)}\right).\nonumber\\
	&        &
\end{eqnarray}
It is then natural to regularize this expansion with a double scaling limit
in which $t\rightarrow 1$ in the interval $0<t<1$  and $n\rightarrow \infty$ with
\begin{equation}
	\label{ds}
	(t-1) n = \mu = \mathrm{fixed}.
\end{equation}
We sum separately the terms with even $k=2h$ and odd $k=2h+1$ in the first series in~(\ref{sing})
by changing sums in $(k,j)$ into sums in $(s,j)$ with $s=h-j$. The second series in~(\ref{sing})
can be evaluated as a sum of iterated integrals of the geometric series. Finally, in terms of
\begin{equation}
	\label{tau}
	\tau = \frac{\alpha}{\mu},
\end{equation}
we find that
\begin{equation}\label{gene}
-\ln\left| Z_n^{\mathrm{sing}} \right| \sim f(\mu,\tau)+ \sum_{j=2}^{\infty} \mu^{2-2j} \mathcal{F}_j(\tau),
\end{equation}
where
\begin{equation}
f(\mu,\tau)=\frac{\mu^2}{4} \left( 3\,\tau^2-2\,\tau -2\,(1-\tau)^2\ln(1-\tau) \right)+ \frac{1}{12} \ln(1-\tau)
\end{equation}
and
\begin{equation}
	\label{gen}
	\mathcal{F}_j(\tau) = \sum_{s=0}^{\infty} (-1)^{s+1} \chi_{j,s} \tau^s ,
\end{equation}
with
\begin{equation}
	\label{car}
 	\chi_{j,s} = (-1)^s \frac{B_{2j}}{2j(2j-2)} \left({ 2j-3+s \atop s }\right).
\end{equation}
These $\chi_{j,s}$ are precisely the Euler characteristics of the moduli space of Riemann surfaces
of genus $j\geq 2$ and $s\geq 0$ punctures. Therefore  the double scaling expansion~(\ref{gene})
comprises generating functions of the Euler characteristics not only for the unpunctured Riemann
surfaces,  but also for the punctured ones. This feature is a consequence of the dependence of 
the sequence $g_n$  on the parameter $\alpha$.  In particular, if we set $\tau= 0$ we find
\begin{equation}
	\label{expd}
 	-\ln\left| Z_n^{\mathrm{sing}} \right|_{\tau=0}
	\sim  -\sum_{j=2}^{\infty} \mu^{2-2j}\frac{B_{2j}}{2j(2j-2)} ,
 \end{equation}
which reproduces the term of genus $j\geq 2$ of the topological expansion for the free energy of the $c=1$ string.
 \section{Concluding remarks}
In this paper we have discussed generalized large $n$ limits for the Penner model, and have considered the value
$t=1$ only  through a limit $t\rightarrow 1$ of a particular example.
However, the case $t=1$ can be independently  studied  in terms of the asymptotic zero distribution of Laguerre
polynomials in the limit~(\ref{clag}) with $A=-1$. This zero distribution was determined in~\cite{DI11} and also
exhibits a fine structure (see Theorem~1 of~\cite{DI11}). Indeed, the support of the zero density $\rho(z)$
depends on the sequence $\alpha_n$, but now through the parameter
 \begin{equation}
 	\label{ele00}
	m = \lim_{n\rightarrow \infty}[\mathrm{dist}(\alpha_n, \mathbb{Z}_n)]^{1/n},
\end{equation}
where $\mathbb{Z}_n=\{-1,-2,\ldots,-n\}$. As a consequence, there is also in this case a one-parameter family
of large $n$ limits of the Penner model for $t=1$  depending  on the value of $m$.
However, since in general the parameters $l$~(\ref{ele}) and $m$~(\ref{ele00}) are different
(take for instance the sequence $\alpha_n=-(n+1)+(2/3)^n$),  the large $n$ limit of the Penner
model for $t=1$ requires a separate analysis.

Multi-Penner models of the form 
\begin{equation}
	\label{mp}
	W(z) = \sum_{i=1}^k \mu_i \log (z-q_i),
\end{equation}
are relevant to characterize the correlation functions of the $d=2$ conformal
$A_1$ Toda field theory~\cite{DI09}. Therefore, it is interesting to investigate their large $n$ limits~\cite{AL14}.
The simplest case  is
\begin{equation}
	\label{jac}
	W(z) = \mu_1 \log(z-1) + \mu_2 \log(z+1),
\end{equation}
which is closely related to the theory of Jacobi polynomials
\begin{equation}
	\label{ja}
	P^{(\alpha,\beta)}_n(z)
	=
	2^{-n}
	\sum_{k=0}^n \left({n+\alpha \atop n-k}\right)
	\left({n+\beta \atop k}\right) (z-1)^k (z+1)^{n-k}.
\end{equation}

The asymptotic zero distribution of Jacobi polynomials in the large $n$ limit 
\begin{equation}
	\label{ab}
	n\rightarrow \infty,\quad \frac{\alpha_n}{n}\rightarrow A,\quad \frac{\beta_n}{n}\rightarrow B, 
\end{equation}
for $A,B\in \mathbb{R}$ has been determined in~\cite{KU04b} and~\cite{MA05}.
It turns out that, as in the case of Laguerre polynomials, the zero density and its support depend
not only on $A$ and $B$ but also on additional parameters related to the degree of approximation
of the sequences $\alpha_n$ and $\beta_n$ to the set $\mathbb{Z}$ of integers.
In this way the large $n$ limit of the two-Penner matrix model~(\ref{jac}) exhibits a fine structure
similar to the Penner model studied in the present work. We finally mention that the analysis
of these features for general multi-Penner models~(\ref{mp}) would require the characterization
of the asymptotic zero distribution of Stieltjes polynomials~\cite{MA11}.
 \section*{Appendix A}
To compute $\re g(a)$ as a function of $t$ we write
\begin{eqnarray}
	\label{ge5}
	\re g(a)& = & \re \Big(\frac{1}{2\pi\rmi t}\int_{C_l} \log(a-z)\,\frac{r(z)}{z} \rmd z \Big)\nonumber\\
                    &    & {}+ \frac{1}{2\pi  t}\int_a^b \ln(x-a) \frac{\sqrt{(x-a)(b-x)}}{x}\rmd x.
\end{eqnarray}
The first term reduces to $(\ln a)/t$. As for the second term
\begin{eqnarray}
	\label{inti}
	\lefteqn{ \frac{1}{2\pi} \int_a^b \ln(x-a)\,\frac{\sqrt{(x-a)(b-x)}}{x} \rmd x = }
	\nonumber\\
	& & \frac{1}{4} \left( a - b + 2\ln \frac{b+1}{1-a} + (a+b)\ln\frac{b-a}{4} \right).
\end{eqnarray} 
Taking into account that $b=1/a$, from~(\ref{ge4b})--(\ref{inti}) we get
\begin{eqnarray}
	\label{jea}
	 \re g(a) & = & \frac{1}{4 t}
	                       \left[
	                              a - \frac{1}{a} -2\ln 2 \left(a+\frac{1}{a}\right)- \left(a+\frac{1}{a}-2\right)\ln a
	                       \right.\nonumber\\
                      &    & \left.
                                      {}+ \left(a+\frac{1}{a}+2 \right)\ln (1+a)+\left(a+\frac{1}{a}-2 \right)\ln (1-a)
                                \right],
\end{eqnarray} 
so that 
\begin{eqnarray}
	\label{uintab}
	U\big |_{[a,b]} & = & \frac{1}{t}(a+\ln a)-2  \re g(a) \\
	                       & = & \frac{1}{2 t}\left[\left(a+\frac{1}{a}\right)(1+2\log 2) + \left(a+\frac{1}{a}\right)\ln a\right.\nonumber\\
                                &   & \left.{}-\left(a+\frac{1}{a}+2 \right)\ln (1+a)-\left(a+\frac{1}{a}-2 \right)\ln (1-a)\right].
\end{eqnarray} 
Thus, using the relations
\begin{equation}
	a+\frac{1}{a} = 4t-2,
\end{equation}
\begin{equation}
	2 \ln (1+a) -\ln a=\ln t+2\ln 2,
\end{equation}
\begin{equation}
	2\ln(1-a)-\ln a = \ln(t-1)+2\ln 2,
\end{equation}
it follows at once that
\begin{equation}
	\label{val000}
	U\big |_{[a,b]} = \Big(2-\frac{1}{t} \Big)-\ln t-\Big(1-\frac{1}{t} \Big)\ln (t-1).
\end{equation}

Let us now consider 
\begin{eqnarray}
	\label{efe0}
	\lefteqn{\int_{\gamma_l}V(z) \rho(z)  |\rmd z|}\nonumber\\
	& = &
	\int_{\gamma_l} \ln |z| \rho(z)  |\rmd z| + \frac{1}{2\pi t} \int_a^b \sqrt{(x-a)(b-x)}\rmd x \\
	& = & \re g(0) + \frac{(b-a)^2}{16 t} \\
	& = & \re g(0) + t-1.
\end{eqnarray}
The computation of $\re g(0)$ requires  a more delicate treatment. We observe that
\begin{equation}
	\label{rge1}
	\re g(z) = \int_{\gamma_l} \ln |z-z'| \rho(z') |\rmd z'|
\end{equation}
is a continuous function on $\mathbb{C}$ which is   harmonic in $\mathbb{C}\setminus \gamma_l$.
We already know that~(\ref{ge3b}) is the form of this function in the unbounded connected component
$D_{\infty}$  of $\mathbb{C}\setminus (C_l\cup [a,\infty))$. To determine $\re g(z)$ in  the bounded connected
component $D_0$  of $\mathbb{C}\setminus (C_l\cup [a,\infty))$ we introduce the function
\begin{equation}
	\label{efe11}
 	f(z) = \int_a^z \frac{r(z')+1}{z'} \rmd z' + \ln a,\quad z\in \mathbb{C}\setminus [a,\infty),
\end{equation}
which is analytic on the complex plane with a cut along $[a,\infty)$. We have that
 \begin{equation}
 	\label{efe2}
 	\ln|z| = t \ln l + \re f(z),\quad z\in C_l.
 \end{equation}
Then from~(\ref{ge3b}) we obtain
\begin{equation}
	\label{efe3}
        \re g(z) = \frac{1}{2t}\Big(x-a-\ln a + \re f(z)\Big)+  \re g(a)+\ln l, \quad z\in \gamma_l.
\end{equation}
Since the right-hand side of this equation is a harmonic function in $ \mathbb{C}\setminus [a,\infty)$,
it follows that the same expression for $\re g(z)$ holds in $D_0$. As a consequence we find
\begin{eqnarray}
	\label{efe4}
	\re g(0) & = & \frac{1}{2t}\Big(-a-\ln a + \re f(0)\Big)+  \re g(a)+\ln l,\\
	            & = & -\frac{1}{2}U\big |_{[a,b]} + \frac{1}{2t} \re f(0)+\ln l.
\end{eqnarray}
Moreover, we have that
\begin{eqnarray}
	\label{efe5}
	\re f(0) & = & \ln a+ \int_0^a\frac{\sqrt{(a-x)(b-x)}-1}{x}\rmd x \\
	           & = & 2\ln 2-1 + \ln a+ \Big[ \frac{1}{2}\Big( a+\frac{1}{a}\Big)-1 \Big]\ln(1-a)\nonumber\\
	           &    &  {}-\Big[ \frac{1}{2}\Big( a+\frac{1}{a}\Big)+1 \Big]\ln(1+a) \\
                   & = & -1+(t-1) \ln(t-1)-t \ln t.
\end{eqnarray}
Substituting~(\ref{val000}) and~(\ref{efe5})  into~(\ref{efe4}) then~(\ref{efe0}) implies that
\begin{equation}
	\label{intv}
 	\int_{\gamma_l}V(z) \rho(z)  |\rmd z| = \ln l + t - 2 + \Big(1- \frac{1}{t}\Big)  \ln (t-1).
\end{equation}
Hence, from~(\ref{efe01}), (\ref{val000}) and~(\ref{intv}) we finally get
\begin{equation}
	\mathcal{E} = -\frac{1}{2}\ln t+\frac{3}{2}\Big(\frac{t-1}{t}\Big) -\frac{1}{2}\Big(\frac{t-1}{t}\Big)^2\ln(t-1).
\end{equation} 
\section*{Acknowledgments}
We thank Prof.~A.~Mart\'{\i}nez  Finkelshtein  for calling our attention to many nice  results on zero asymptotics
of Laguerre polynomials. The financial support of the Ministerio de Ciencia e Innovaci\'on
under project FIS2011-22566 is gratefully acknowledged.

\end{document}